\documentstyle[epsf,psfig,12pt]{article}
\def\hmpc{\rm \,h^{-1}\,Mpc}
\def\etal{{\it et al.}\ }
\def\x{$\xi(r)$\ }

\def\kms{\,{\rm km\,s^{-1}}}

\def\ro{r_\circ}
\def\n_med{{\left<n\right>}}

\def\begc{\begin{center} }
\def\endc{\end{center} } 
\def\begf{\begin{figure} }
\def\endf{\end{figure} }
\def\IRAS{{\it IRAS}}

\def\fig #1, #2, #3 {
\smallskip
\centerline{\psfig{figure=#1,height=#2 in,width=#3 in}}
}

\title{IS THE UNIVERSE HOMOGENEOUS?\\ 
(On Large Scales)
\footnote{Accepted for publication in {\it New Astronomy} [2, 517 (1997)].  
Full html version
available at 
{\tt http://pigeon.elsevier.nl/journals/newast/}.  Original colour figures
available at {\tt ftp://antares.merate.mi.astro.it/pub/guzzo/NA}.
}}

\author{Luigi Guzzo
\\ \\
Osservatorio Astronomico di Brera \\
I-22055 Merate (LC), Italy. \\
}

\date{}

\begin{document}

\maketitle

\begin{abstract}

	I critically discuss in a pedagogical and phenomenological way 
	a few crucial tests challenging the recent claims by 
	Pietronero and collaborators that there is no evidence from available 
	galaxy catalogues that the Universe is actually homogeneous above 
	a certain scale.  In a series of papers, these authors assert that 
	observations are consistent with a fractal distribution of objects 
	extending to the limit of the present data.  I show that while 
	galaxies are indeed clustered in a scale--free (fractal) way
	on small and intermediate scales, this behaviour does not 
	continue indefinitely.  
	Although the specific wavelength at which the galaxy distribution
	apparently turns to homogeneity is dangerously close to the size of the
	largest samples presently available, there are serious 
	hints suggesting that this turnover is real and that its effects are
	detected in the behaviour of statistical estimators.  The
	most recent claims of a continuing fractal hierarchy up to 
	scales of several hundreds Megaparsecs seem to be abscribable to 
	the use of incomplete samples or to an improper treatment 
	of otherwise high--quality data sets.
	The fractal perspective, nevertheless, represents a fruitful way to
	look at the clustering properties of galaxies, when properly coupled 
	to the traditional gravitational instability scenario. In the last
	part of this paper I will try to clarify, at a very simple level,
	some of the confusion existing on the actual 
	scaling properties of the galaxy distribution, and discuss how 
	these can provide hints on the evolution of the large--scale 
	structure of the Universe.

\end{abstract}

\noindent{\it PACS Codes: 98.80.-k}, {\it Keywords: Cosmology: large--scale
structure of universe --- cosmology: observations --- cosmology: theory}

\section{Introduction}

``The Universe is spatially homogeneous and isotropic'', states the
Cosmological Principle, the pillar on which the standard cosmological model 
is built.  It is this assumption that allows us to treat the spatial
hypersphere of our Universe (i.e. the three--dimensional space we live in),
as a maximally symmetric subspace of the whole of space--time, and derive
what is known as the Robertson--Walker metric.  All we do in our
everyday cosmology work is based upon this, with the 
further ingredient of an equation of state for the material content of 
the Universe  (see e.g. Weinberg 1972).   
Distances, luminosities, sizes: they all depend on this initial 
assumption.  For this reason, considerable efforts were 
spent in the past decades to test observationally the validity 
of the Cosmological Principle, as described, e.g., in Peebles (1993).
Due to the lack of distance measurements for most of the galaxies, 
the prime test adopted was that of counting objects as a function 
of apparent flux, and still number counts are widely used as a powerful 
probe of the distant/past Universe (see e.g. Ellis 1997).
The situation changed in the seventies, when the industry of galaxy 
redshift measurements started\footnote{See Rood 1988 for a colorful historical 
review of the pioneering years of redshift surveys.}, allowing the 
reconstruction of the
true tridimensional galaxy distribution over large areas of the
sky.  To everybody's surprise, since then (and until very recently), 
any new 3D map continued to show highly structured patterns, with sizes
comparable to those of the volumes surveyed.  The question became, therefore, 
on which scales did the expected homogeneity start, or better, 
was there really any evidence of the assumed homogeneity?   
This specific observation
that larger and larger surveys seemed to show larger and larger 
structures naturally suggested a similarity with the mathematical objects
called {\it fractals}, that were becoming commonplace in the description 
of other physical phenomena, in particular of what was known as 
{\it nonlinear dynamics}.
Several pages of B.~Mandelbrot book {\it The Fractal Geometry of Nature},
are dedicated to this property of the observed galaxy distribution.  
In fact, the elegant idea of a continuous fractal hierarchy of cosmic
structures goes back to the beginning of the century (Charlier 1908), 
but only in recent years the actual data
could be used as direct motivation for such models.  Further 
support to the fractal view was added
by the first estimates of the galaxy correlation function, whose power--law
shape quantitatively indicated a scale--free character typical of fractal 
sets (Peebles 1980).

Given the importance of the general issue, one can understand the reasons for
the considerable -- and very often controversial -- debate that arised 
on this subject, stimulated in particular by the work of Pietronero and 
collaborators.  In particular, the first paper by this latter author on 
this specific subject (Pietronero 1987) 
is a good reference to grasp the spirit of the controversy, and can be 
taken as the manifesto of the ``No--evidence--for--homogeneity'' line 
of thought.  Several debates have 
been organized in the last couple of years in the attempt to settle 
this controversy, the most notable one being probably that of the 
``Critical Dialogues in Cosmology'' conference, 
held in Princeton in June 1996, opposing L.~Pietronero (Pietronero \etal\ 
1997, P97 hereafter), and M.~Davis (Davis 1997).

This paper has its origin in a similar debate, held in Frascati (Rome)
in December 1996 during the fifth Italian National Cosmology Meeting, in which 
I was asked to defend the thesis that the Universe is indeed homogeneous 
on large scales, against F.~Sylos--Labini.  Here I present in some 
detail the conclusions I have reached by looking at essentially two
simple tests based on redshift data.  This, therefore, is not conceived
as a thorough review on the subject -- although it often summarizes and
uses previous results with a pedagogical style -- but aims at showing 
through a couple of very specific observations that our faith in the 
Cosmological 
Principle is still well placed.  Since very often the controversies on
this and other subjects have arised by a lack of communication produced
by the different jargons used in different fields of physics, the style
is deliberately phenomenological and aims at being understendable to 
an audience possibly larger than just the astronomical community.

I decided to base my defense essentially on: a) the scaling of the 
galaxy--galaxy correlation length $\ro$ with the sample size; 
b) the behaviour of counts of objects within volumes of increasing radius 
centered on our own galaxy.  These are the two main arguments used by 
Pietronero and collaborators to argue that in fact the data are consistent
with an unlimited fractal with dimension $D\sim 2$ (e.g. P97).
The topics I chose are also partly complementary to those discussed by 
Davis at the Princeton meeting (Davis 1997).  More
discussion on the subject can be also found in Peebles (1993).

Since in this controversy I feel like playing the role of the advocate, 
this paper can be seen as a sort of harangue, whose aim will be to 
convince the jury that when the best optical redshift surveys are properly
analyzed, the data are more consistent with a Universe that becomes 
homogeneous above a scale comprised between 100 and 200
$\hmpc$, rather than with an unlimited fractal distribution.  Trying to be 
a good lawyer, in the final section I will try to clarify some 
of the confusion existing on the actual scaling properties of the galaxy
distribution, {\it in the range where there is indeed a fractal--like 
scaling,} and discuss how these can provide some hints on the origin of the 
large--scale structure of the Universe.

\section{Fair Samples}

The first point that needs to be clarified before entering the core
of the problem, is that any investigation that aims at checking
for large--scale homogeneity cannot be based on catalogues that are
intrinsically highly incomplete.  I refer in particular to the LEDA 
database (Paturel \etal\ 1995), on which some of the claims for a neverending
fractal structure have recently been made (Di Nella \etal\ 1996, P97).  
This catalogue  is a completely heterogenous 
compilation of literature redshifts, which can be very useful for a number
of other applications, but whose intrinsic spatial inhomogeneities make it 
useless for performing an unbiased clustering
analysis.  These incompletenesses are addressed in a rather naive way in the
quoted work (where they are treated as {\it spatially
random}, while they are {\it patchy}), and are most
probably the origin of the enormous correlation length, $r_\circ \simeq 45 
\hmpc$, estimated for the largest sample extracted from this catalogue.   
Since nowadays we have much better data sets available for testing our
theoretical expectations, in this discussion I will not consider the results 
coming from LEDA as having any meaning whatsoever.  

\section{Scaling of the Correlation Length: $\ro = f(R_s)$?}

\subsection{Basics}
\label{basics}

Let us remind the basic properties of fractal structures, and
their relation to the standard correlation function used in 
cosmology \x (see e.g. Pietronero 1987 and Provenzale 1991 for
a more comprehensive discussion).
A fractal distribution of points is characterized by a specific 
{\it scaling} relation.  This scaling law can be by itself taken
as a definition of fractal: the number of objects counted in 
spheres of radius $r$ around a randomly chosen object in the set
must scale as
\begin{equation}
N(r) \propto  r^{D}\,\,\,\,\, ,
\label{n_z}
\end{equation}
where $D$ is the {\it fractal dimension} (or more correctly, the fractal
{\it correlation} dimension, see Provenzale 1991).  Analogously, 
the density within the same sphere will scale as
\begin{equation}
n(r) \propto  r^{D-3}\,\,\,\,\, .
\end{equation}
For our purposes, it is useful to consider the expectation value of 
the density measured within shells of width $dr$ at separation $r$ from an 
object in the set.  This is called the
{\it conditional density} $\Gamma(r)$, after Pietronero (1987), 
and for a fractal set will scale in the same way, 
\begin{equation}
\Gamma(r) = A \cdot  r^{D-3}\,\,\,\,\, ,
\label{gamma}
\end{equation}
where $A$ is an amplitude that is a constant for a given fractal set.
This function is useful, because it allows us to connect the fractal
description to the standard two--point correlation function \x 
commonly used in cosmology, i.e. to the excess probability of
finding an object within the same shell of width $dr$ at separation $r$ 
from another taken at random (see Peebles 1980 for an accurate definition).  
If $\n_med$ is the mean density of the sample (we shall come back
in a moment to explain what this means for a fractal), then it
can be shown that
\begin{equation}
1+\xi(r) = {\Gamma(r) \over \n_med} \propto  r^{D-3}\,\,\,\,\, .
\label{xipiu}
\end{equation}
In the strong clustering regime $\xi(r)\gg 1$, and therefore we
recover the classic result discussed in Peebles (1980; 1993), that a 
power--law correlation function for galaxies, $\xi(r) = (r/r_\circ)^{-\gamma}$,
implies a scale--free, fractal clustering with dimension $D=3-\gamma$.
This formally justifies the common wisdom that with a power--law
correlation function the structures have a fractal character and look 
the same no matter the scale at which they are viewed.  
Note, however, that in the weak clustering regime only $1+\xi(r)$, 
and not $\xi(r)$, will be able to show the existence of a fractal--like 
scaling.   I will come back to this in \S~\ref{true-scaling}.

The discussion so far is valid for the case of a pure fractal distribution, 
or more in general within a range of scales where a clear scaling law can be 
identified (as for a fractal range limited by an upper transition
to homogeneity).
In this situation the mean density $\n_med$ is an ill--defined
quantity, and depends directly on the sample size within which 
it is measured.  
In fact, integrating eq.~\ref{gamma} one can see that
for a spherical sample with radius $R_s$
({\it centered on an object}, a point not always appreciated),
the mean density is
\begin{equation}
\n_med = {3 A \over D} \cdot R_s^{D-3}\,\,\,\,\, ,
\label{meandens}
\end{equation}
i.e. decreasing as a power law of the sample radius $R_s$.
Correspondingly, the two--point correlation function is
\begin{equation}
\xi(r) = {\Gamma(r) \over \n_med} - 1 = {D \over 3} \cdot \left({ r 
\over R_s}\right)^{D-3} - 1\,\,\,\,\, ,
\end{equation}
with a correlation length
\begin{equation}
r_\circ = \left({6 \over D}\right)^{1 \over D-3} \cdot R_s\,\,\,\,\, .
\label{r0}
\end{equation}

We have therefore a specific prediction: if the galaxy distribution has
a pure fractal character, with a well defined fractal correlation 
dimension $D$,
then the correlation length $r_\circ$ is a linear function of the sample 
radius $R_s$, with a coefficient that depends on the fractal dimension.

\subsection{What Do the Data Say?} \label{data-sec}

The claim by Pietronero and coworkers is that the behaviour described 
by eq.~\ref{r0} is actually observed in the data.
We know that galaxy clustering is well described by a power--law 
correlation function at small separations, at least for $r < 5 - 10 \hmpc$.
This certainly implies that for sufficiently small samples the correlation 
length will not be a stable quantity.  
This should not be a tremendous surprise even for the not--fractally--trained 
cosmologist: it is a natural consequence of clustering independently from
the way we describe it.  A slightly different way of expressing the same 
concept, is to say that if one measures the mean
density within a sphere of radius $R_s$ centered on a galaxy (so as
our redshift surveys are centered on our own galaxy), the expectation
value of this quantity, $\bar n$,  will be given by
\begin{equation}
\bar n = {3\over R_s^3}{\int_0^{R_s} \n_med \left[ 1+\xi(r)\right] 
r^2 dr}\,\,\,\, .
\label{n-measured}
\end{equation}
Since \x is positive for $r<30-60 \hmpc$, for values of $R_s$ comparable
to these inevitably $\bar n$ will be larger than the asymptotic cosmic
reference value $\n_med$.  In other words, for the
simple reason that galaxies are  clustered, the mean density as seen 
by a generic galaxy will tend to be systematically overestimated (note that
this is true for the {\it expectation value} and therefore for an ensemble
average: there will be fluctuations
around this, so that some observers could well see an opposite trend.)
Increasing the sample size $R_s$, $\bar n$ will converge asymptotically 
to $\n_med$, in a way that depends on the shape of the {\it true} 
two--point correlation function \x.  
Consequently, in turn, any estimate of \x will tend be underestimated
(i.e. overnormalized) for small samples, its amplitude will grow with
the sample size and the correlation length $\ro$ will be a
function of the sample radius $R_s$, in essentially the same way as
discussed in the previous section.  This was in fact observed {\it 
for small samples} at the end of the eighties (e.g. Einasto \etal\ 
1986), so there is little doubt that this is a real, 
consistent observational fact.   The crucial point to be asked is however 
the following: is the observed 
relation between $\ro$ and $R_s$ linear over a significant range of scales,
or does it asymptotically converge to a stable value?  
Coleman \etal\ (1988) claimed that such relation was indeed linear,
with no evidence for convergency.
However, in order to sample large enough volumes,
in their study of the CfA1 sample they mixed up the clustering properties of 
galaxies with rather different intrinsic luminosity.  In this way, they did
not consider an extra variable (i.e. luminosity), upon which clustering
does, albeit weakly, depend (see e.g. Iovino \etal\ 1993, and more 
recently Guzzo \etal\ 1997).   
Davis \etal\ (1988), for 
example, showed that the functional dependence, at least for \IRAS\ galaxies,
was more like a square root of $R_s$, a possible indication that the fractal 
range had some sort of upper cutoff to homogeneity inducing convergency 
of the mean density.  These results gave rise to
strong controversies related both to the samples used (CfA1 vs. \IRAS) and 
to the use of statistical estimators in the computation of \x.
Rather than going back to the details of the controversy [that can be found 
in Davis (1997) and P97], I prefer to turn to the new
optical redshift data, that have become available since then and see
what they yield in this respect.

In Table~\ref{tab-r0} I have listed as reference the general properties of the
three deepest wide--angle redshift surveys of optically--selected galaxies 
available to date, the Stromlo/APM survey,
the ESO Slice Project (ESP) survey, and the Las Campanas Redshift Survey 
(LCRS).
Redshift surveys typically do not sample a spherical volume, so that the
``sample radius'' has to be defined appropriately.  In this case, 
a reasonable definition of $R_s$ is the radius of the maximum sphere 
which can be contained within 
the survey volume (e.g. P97), that is approximately 
expressed as $R_s =  d \,{\sin({\beta/ 2}) /(1 + \sin (
{\beta/ 2}))}$, 
where $d$ is the survey depth and $\beta$ is the angular diameter of the
largest circle that can be fully contained within the survey area on the
sky.  
\begin{table*}
\begin{center}
\begin{tabular}{lccrcc}  \hline\hline 
Survey & $d$  & $\beta$ & $R_s$  & $\ro$ $^{(predicted)}$&
$\ro$ $^{(observed)}$ \\ 
\hline
ESP 		& $\sim 600$	&  $1^{\circ}$ &  5 &  1.7 & $4.50^{+0.22}_{-0.25}$\\
LCRS 		& $\sim 400$	& $10^{\circ}$ & 32 &  11  & $5.0\pm 0.1$\\
Stromlo/APM 	& $\sim 200$	& $90^{\circ}$ & 83 &  28  & $5.1\pm 0.2$\\
\hline
\end{tabular}
\end{center}
\caption{The behaviour of the correlation length $r_\circ$ for some of the 
most representative modern galaxy redshift surveys, ESP
(Vettolani \etal\ 1997; Bartlett \etal\ 1997), Las Campanas (Lin, 1995), 
and APM/Stromlo (Loveday
\etal\ 1995), compared to the predictions of the $D=2$ model.  All estimates 
of $\ro$ are true correlation lengths in {\it real} space.  $d$ is the 
effective depth of the surveys.  All measures of distance are expressed in 
$\hmpc$.
}
\label{tab-r0}
\end{table*}                                                                   

For the three surveys, Table~\ref{tab-r0} lists, in particular,
the ``sample radius'' 
$R_s$ and the corresponding expected value of $\ro$ for $D=2$, if 
eq.~\ref{r0} holds.  The last column gives the actual observed 
values of $\ro$.  What the data consistently show is that the observed
correlation lengths are off by tens of standard deviations 
from the values predicted by the simple $D=2$ fractal model.  The
result would be even worse using $D=1.2$ (i.e. extrapolating the 
fractal dimension actually observed on small scales -- 
see \S~\ref{true-scaling}), 
that would predict larger $r_\circ$'s.
I stress that these are true {\it real--space} values of the correlation
length, and are performed on samples that include a similar range of
absolute magnitudes.  These aspects are usually confusingly mixed up 
in the comparisons presented, e.g., in P97.  In fact, 
redshift--space correlation functions are characterized by a correlation
length $s_\circ$ that is typically about 15\% higher than the corresponding
$\ro$, due to the combined effect of small--scale and large--scale 
distortions produced by peculiar velocities.  On the other hand, 
volume--limited samples that include only galaxies brighter than 
$M\sim M^*$ yield real--space correlation lengths $\ro \ge 7 \hmpc$
(see e.g. Guzzo \etal\ 1997).

The stability of $\ro$ in Table~\ref{tab-r0} is indeed remarkable, 
and cannot be justified as a consequence of the technique used to 
estimate $\xi(r)$, as often claimed (e.g. P97), given the very different
geometry of the three surveys.  This issue is discussed in a bit more detail 
in Appendix A.

\section{Growth of Galaxy Numbers: $N(r)\propto r^D$ ?}

The other test that I have decided to discuss is the behaviour of
galaxy spatial counts within growing volumes centered on our own
preferred location.  Pietronero and collaborators claim that also this 
statistics does indeed agree with the behaviour expected for an 
unlimited fractal distribution (eq.~\ref{n_z}) for a number of 
redshift surveys [more precisely, in P97 it is discussed the equivalent 
behaviour of the density within growing volumes, i.e. of the quantity
$N(r)/r^3$].  Here I present my version of the story.  First, I have considered 
one of the redshift surveys of the previous generation, the Perseus--Pisces
redshift survey (PP, Giovanelli \& Haynes 1991), which is one of those used
by P97 for their analysis.  
\begf
\fig 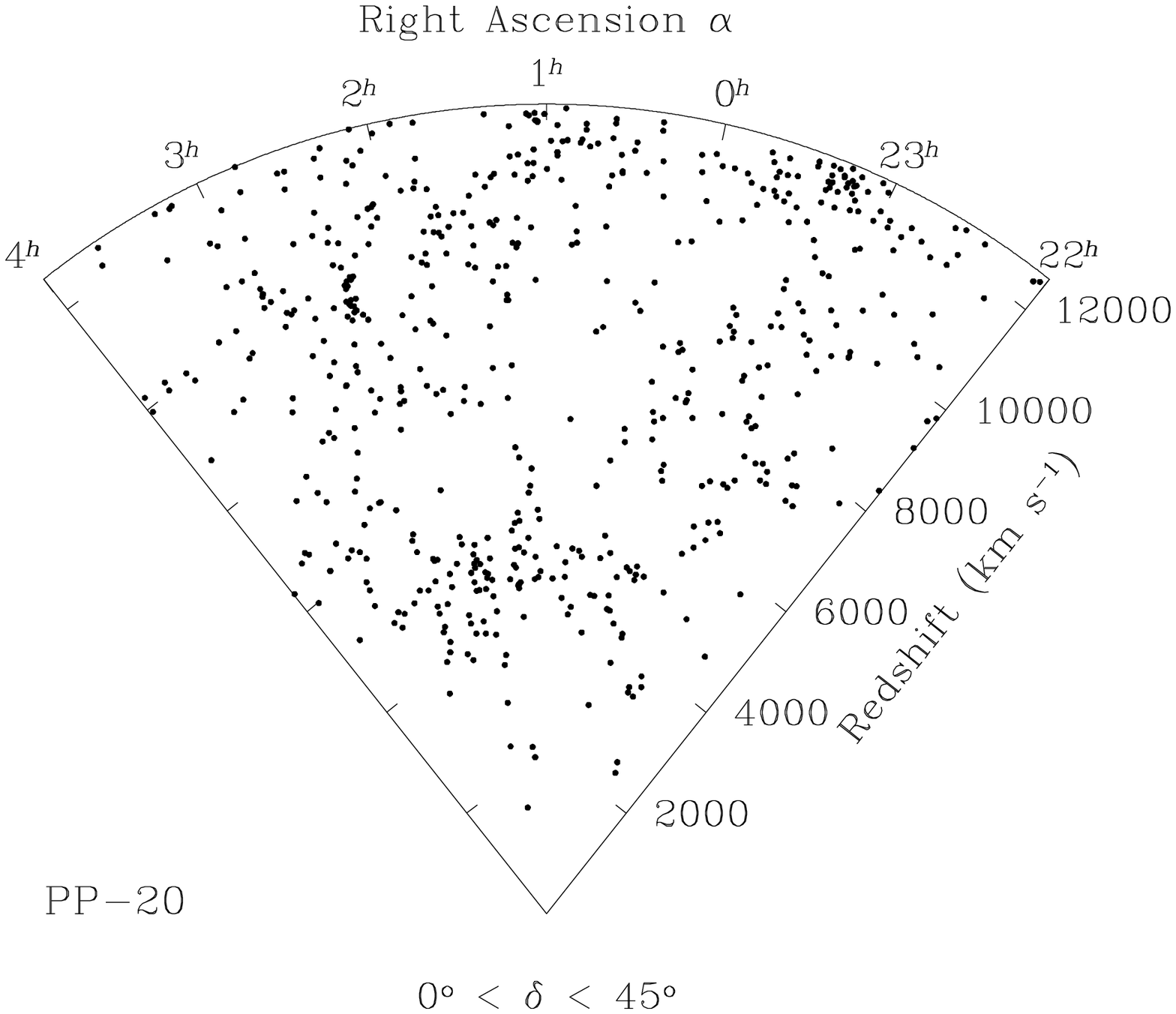, 4, 4
\caption{Galaxy distribution in the $M\le -20$ volume--limited sample of the 
Perseus--Pisces redshift survey used in the $N(r)$ analysis (from
Guzzo \etal\ 1997).}
\label{cone_pp20}
\endf
The cone diagram of Figure~\ref{cone_pp20} displays the distribution of the
galaxies contained within a
volume--limited subsample\footnote{Throughout the paper, I write the Hubble 
constant as $H_\circ = 100\kms$ Mpc$^{-1}$ and compute absolute magnitudes
assuming $h=1$.} with $M\le-20$ and $cz \le 12600\kms$, basically
the deepest volume--limited sample one can construct from the survey.
The reason why I show this cone diagram is to remark what I mentioned in
the Introduction, i.e. how until very recently all redshift surveys continued 
to reveal structures with size comparable to the
survey depth. (Another example is the CfA2 survey, that, equivalently, is dominated 
by the so--called Great Wall, see e.g. Geller \& Huchra 1989).
Figure~\ref{Nz_pp20} (filled circles), shows the result of counting the
number of objects as a function of distance, i.e. the integral counts within 
volumes of radius $cz$ (in $\kms$) for this sample.  The red
line gives the expected behaviour for a fractal distribution with $D=2$,
while the blue line describes the homogeneous case (i.e. $D=3$).   The two 
curves
are normalized as to reproduce the total number of galaxies in the sample.
Taken at face value, this plot seems to indicate that after an initial 
wiggling around the homogeneous curve, 
above $cz\sim 6000 \kms$ the counts rise closer to the fractal curve,
apparently confirming the findings of Pietronero and collaborators.  Reality
is a bit more complicated than this plot alone shows, however.  
\begf
\smallskip
\centerline{\psfig{figure=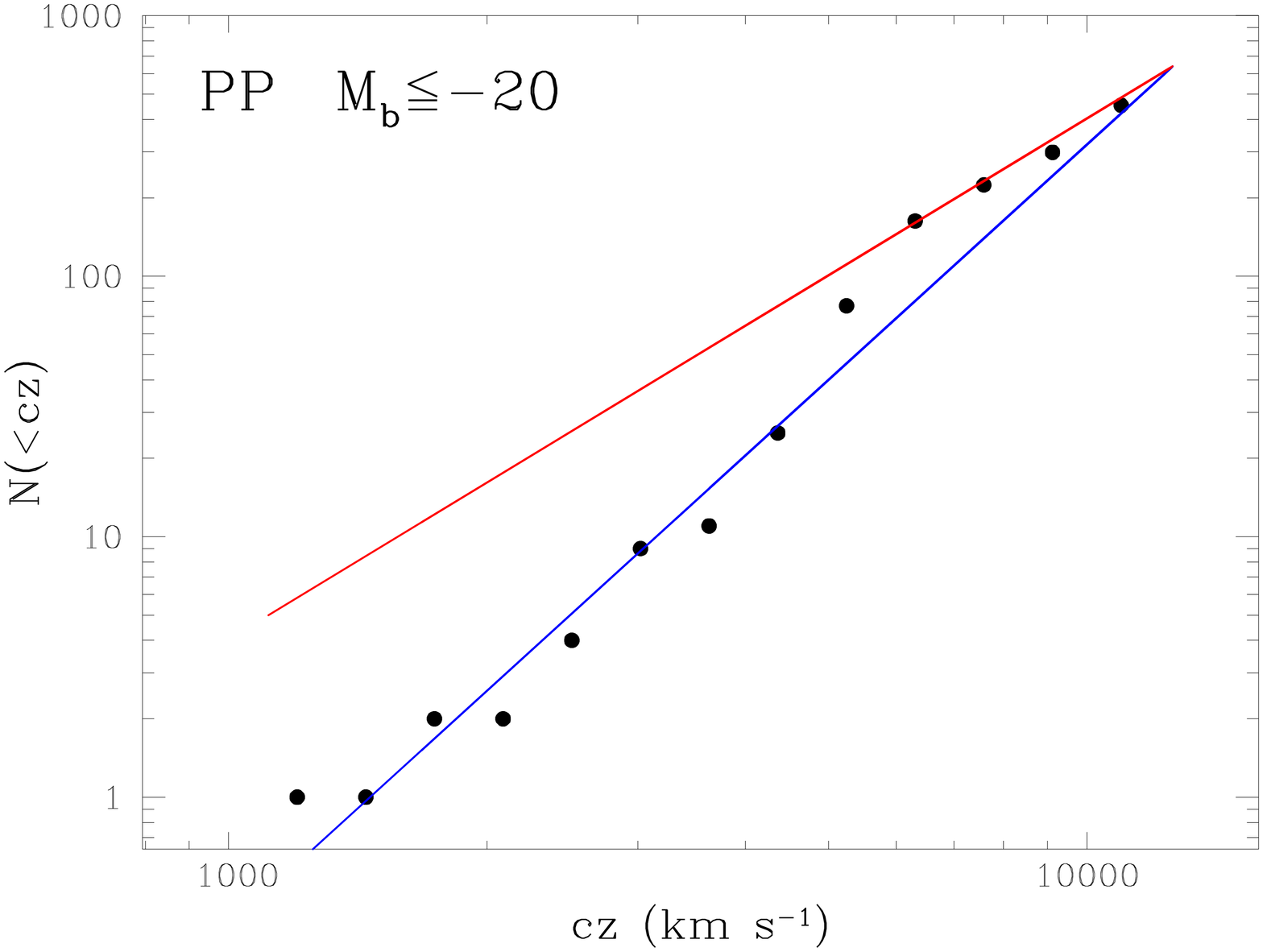,width=4.5in,angle=270}}
\caption{Number of galaxies within volumes of increasing radius in 
the volume--limited ($M\le -20$) subsample of the Perseus--Pisces
redshift survey of Figure~~\ref{cone_pp20}  (filled circles).  
For comparison, the red line gives the relation
expected for a fractal distribution with $D=2$, while the blue 
line refers to a homogeneous model.  Apparently, the observed counts
seem to stabilize around the fractal curve at large radii. (This being a
cumulative distribution, points are not independent and error bars 
would have no meaning.  See Figure~\ref{hist_pp20} for an indication
of the true amount of fluctuation around the differential histogram).}
\label{Nz_pp20}
\endf
Let us 
consider the {\it differential} redshift distribution of the galaxies
for the same sample, shown in Figure~\ref{hist_pp20}.   The solid histogram
in the left panel clearly shows the preponderancy of the Perseus--Pisces 
supercluster at $cz \sim 5500\kms$, while again the two curves in colour
show the expectation values of the counts in the $D=2$ and homogeneous
cases.  One can notice that the boost observed in the integral counts 
of Figure~\ref{Nz_pp20} starts where the strong positive
contribution of the supercluster enters the sampled volume.  The right 
panel shows the ratio between the observed counts and those predicted
in the two scenarios, (the curve whose mean trend is closer to 1 will
be the one which better describes the data).  The red line (fractal case),
shows a monotonic increase with scale, indicating that it predicts 
on the average less 
objects at larger and larger radii than are actually observed.  The 
homogeneous model describes the data better, fluctuating around a 
substantially flat mean, although the differences between the two models for 
$cz>4000 \kms$ are very small and not significant.   
Intrinsic fluctuations due to
clustering are still very strong on samples this size, and also when
averaged over the large solid angle of the PP survey ($\sim 1\, sr$),
they are much stronger
than the Poissonian fluctuations within the bins (indicated by the error
bars).  Comparing the integral and differential histograms, one should 
probably conclude that samples of this depth are simply not large enough 
for using $N(r)$ as a discriminator of the global geometry.
\begf
\centerline{\psfig{figure=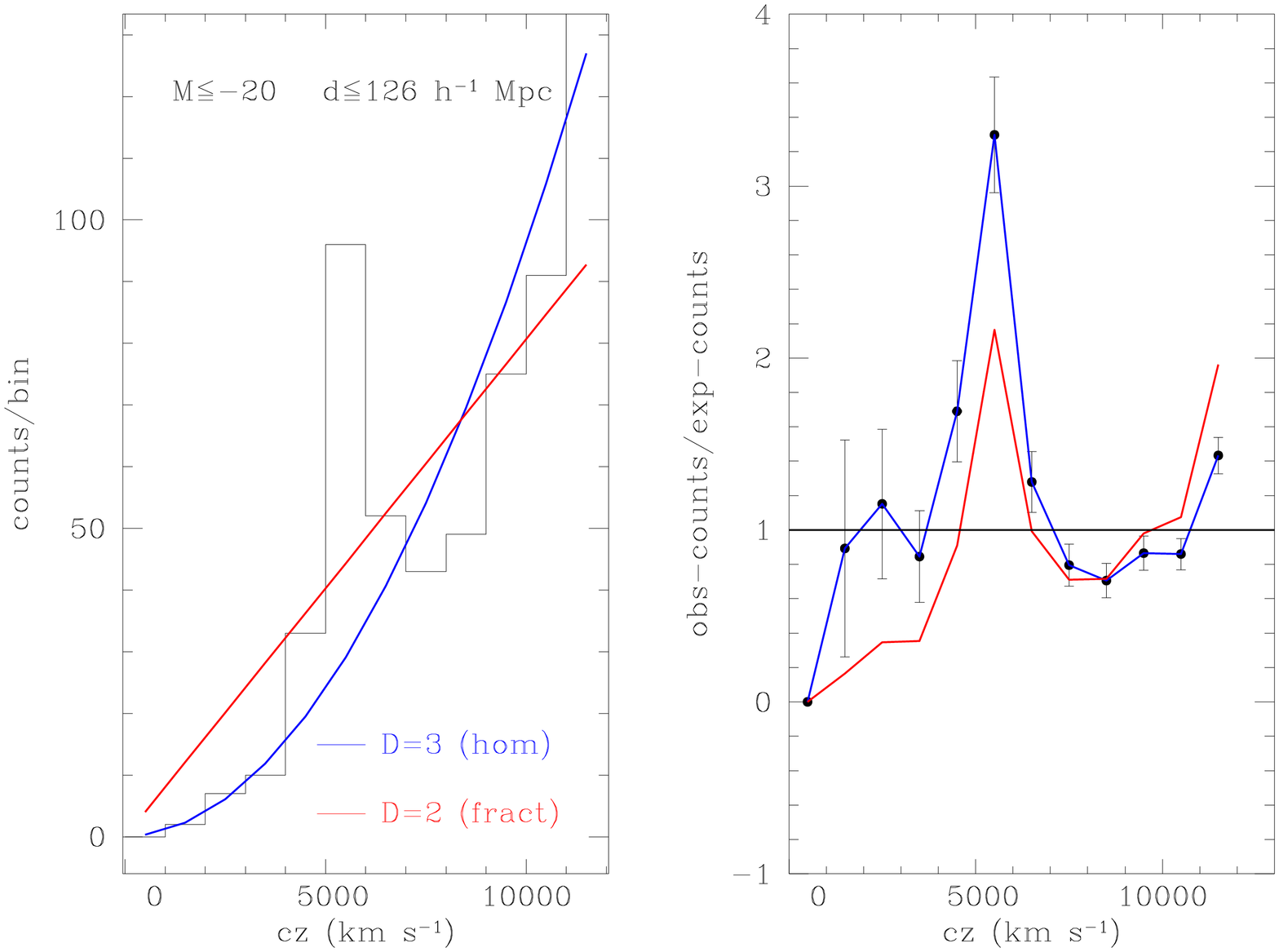,width=4.5in,angle=270}}
\caption{Histogram of the galaxy distribution as a function of distance
from the observer (left panel) for the PP survey subsample.  As in the 
previous figure, the red and
blue lines give the expected behaviour for a $D=2$ fractal and a homogeneous
distribution, respectively.  In the right panel, the observed differential
counts have been normalized to those expected for the two models.  The 
strong positive fluctuation around $cz\simeq 5500 \kms$ corresponds to
the Perseus--Pisces supercluster.}
\label{hist_pp20}
\endf

Let us therefore consider again what are presently the state--of--the--art 
for redshift surveys, and
in particular the ESP.   This redshift survey has an
apparent magnitude limit of $b_j=19.4$ and contains radial velocities for 
3342 galaxies, whose distribution in space is shown in Figure~\ref{esp_cone}.  
Note for comparison how the largest structures
in this survey are much smaller than the survey size (what R.~Kirshner
called ``the end of greatness'' when describing the similar LCRS), unlike 
the PP and CfA2 surveys\footnote{This by itself is an observation
that is rather difficult to reconcile with an unlimited fractal 
distribution on all scales.}.   Based on the analysis of a 
preliminary version of this survey, Pietronero and collaborators again
claim that the
behaviour of $N(r)$ is consistent with a $D=2$ fractal.  I have repeated
the same exercise with the whole survey (now completed, Vettolani \etal\ 1997),
 extracting a volume--limited sample 
with $M\le -20$ and $z\le 0.189$, containing a total of 517 galaxies.  
One word of caution is important here: this survey is deep enough so that 
galaxy magnitudes have to be K--corrected before the true rest--frame 
absolute magnitude can be recovered (see Zucca \etal\ 1997 for details on 
the K--correction procedure).
Apparently, this seems to have been neglected in the analysis presented, e.g., in
P97 [see Scaramella \etal\ (1997) for a more detailed investigation of this point, together
with a comparison of the $N(r)$ of Abell clusters].   In Figure~\ref{Nz_esp}
I have plotted the integral counts $N(z)$ obtained for the ESP subsample, compared 
again to
the expectations of the $D=2$ and $D=3$ cases, as in Figure~\ref{Nz_pp20}. 
\begf
\fig 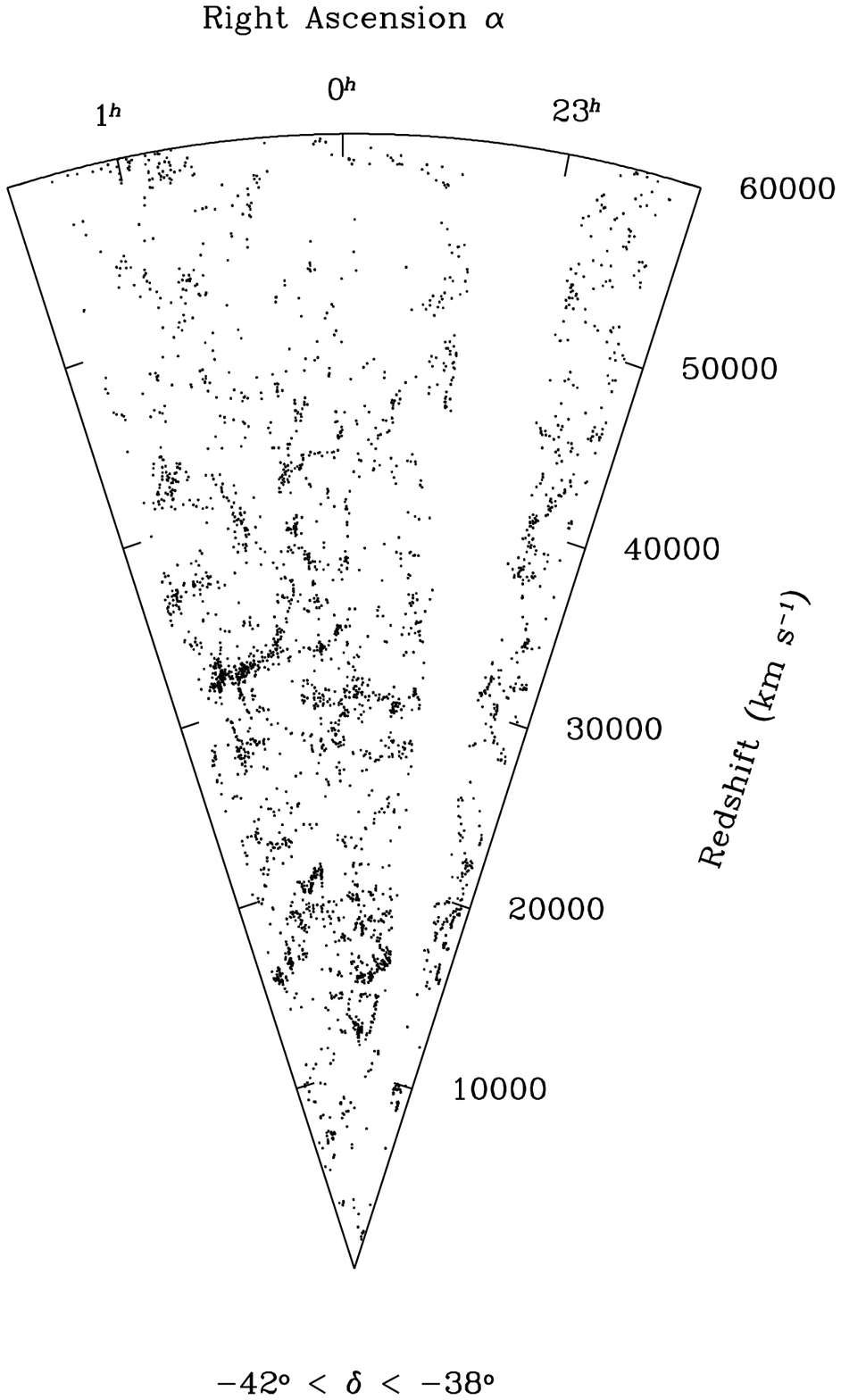, 6, 6
\caption{Galaxy distribution in the ESO Slice Project (ESP) redshift
survey.}
\label{esp_cone}
\endf
\begf
\centerline{\psfig{figure=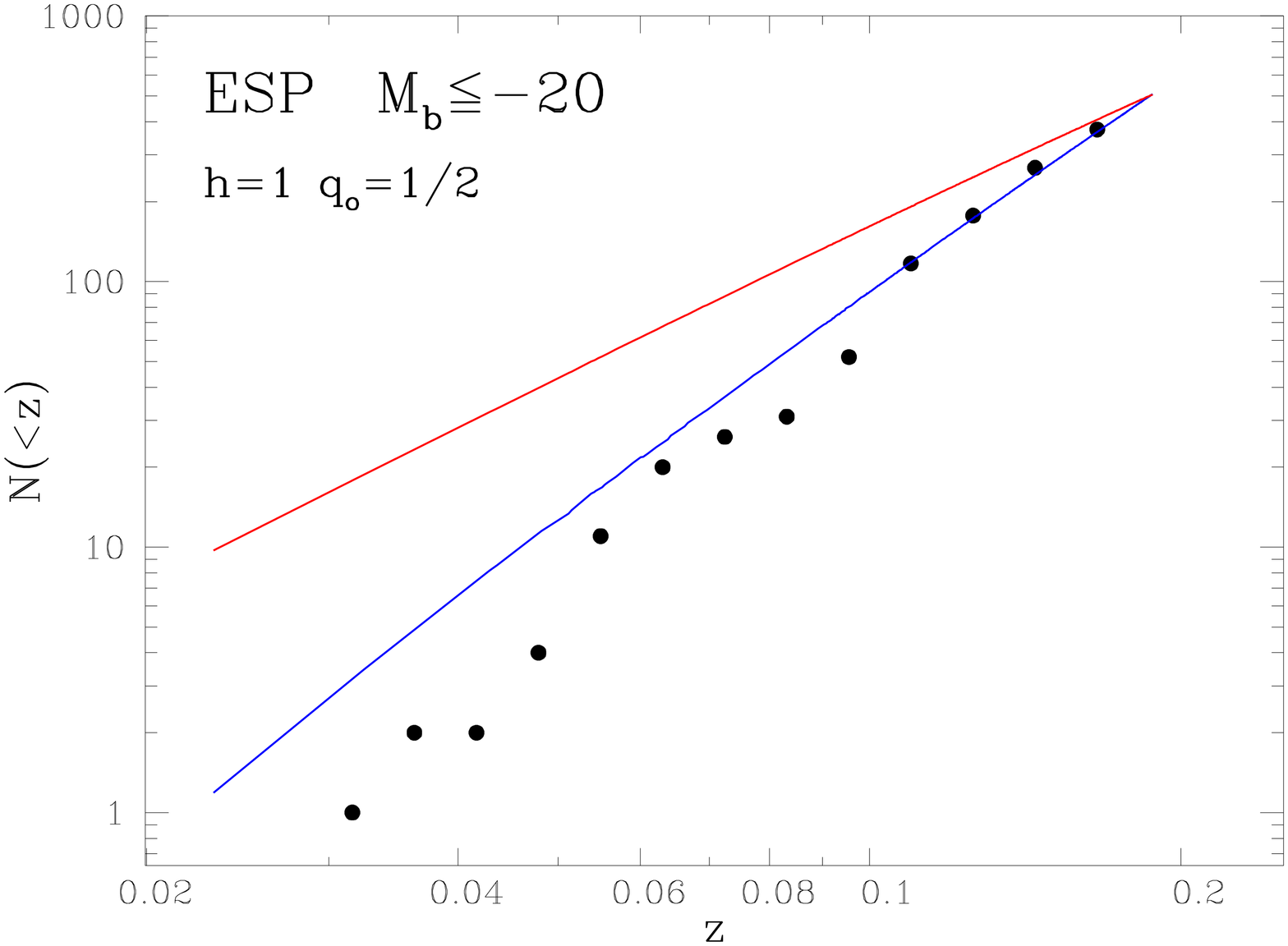,width=4.5in,angle=270}}
\caption{Number of galaxies within volumes of increasing radius in 
a volume--limited ($b_J\le -20$) subsample of the ESP  
redshift survey (filled circles).  The red line gives the relation
expected for a fractal distribution with $D=2$, while the blue 
line refers to a homogeneous model.  Unlike for the PP survey, and 
in contrast with the claims by Pietronero \etal, the observed counts 
on large enough scales seem to stabilize around the homogeneous model.}
\label{Nz_esp}
\endf
The counts 
are now performed in the full cosmological framework, assuming a spatially flat
universe with $\Omega = 1$, as indicated in the figure.  
The fractal (red) curve is computed by assuming that comoving distances
are distributed according to the fractal law, and then calculating the
corresponding redshift in the chosen world model.   I note, however,
that there is an internal inconsistency in this procedure, since the 
Friedmann--Robertson--Walker cosmology {\it assumes} a 
priori homogeneity on large
scales.  Therefore, we would not even be allowed to make this comparison
on scales where the global cosmology starts to be important, if the fractal 
hypothesis were true.  In other words, one would have first to re--found 
a different cosmology, based on a different sort of Cosmological Principle
compatible with the fractal hypothesis, construct the proper distance 
relations, and then finally compare predictions with the observations.
For our purposes, cosmological corrections are still small 
on these scales so that the simple FRW comparison should be meaningful.

As one can note from the figure, the integral counts seem to be very
well in agreement with those expected for a homogeneous Universe on
large scales.  Note the low values of the counts at small radii. This
is due first of all to the small survey volume within these distances
(the ESP survey area is only $1^\circ$ thick in declination), so that 
bright objects
are lost due to shot noise, that cuts off the bright tail of the 
luminosity function.  (This is the same as saying that the volume is not large
enough as to make the integral of the luminosity function over that
volume larger than 1:  since we cannot observe, say, 1/5 or 1/2 of a galaxy,
no galaxy -- at that luminosity -- is observed).  The second effect is a 
true underdensity in the mean distribution of galaxies in the South Galactic
Pole region which seems to extend out to $\sim 140 \hmpc$.  This underdensity
has been shown by Zucca \etal\ (1997) to be the origin of the observed deficit 
of ``bright'' galaxies in the number counts.    
\begf
\centerline{\psfig{figure=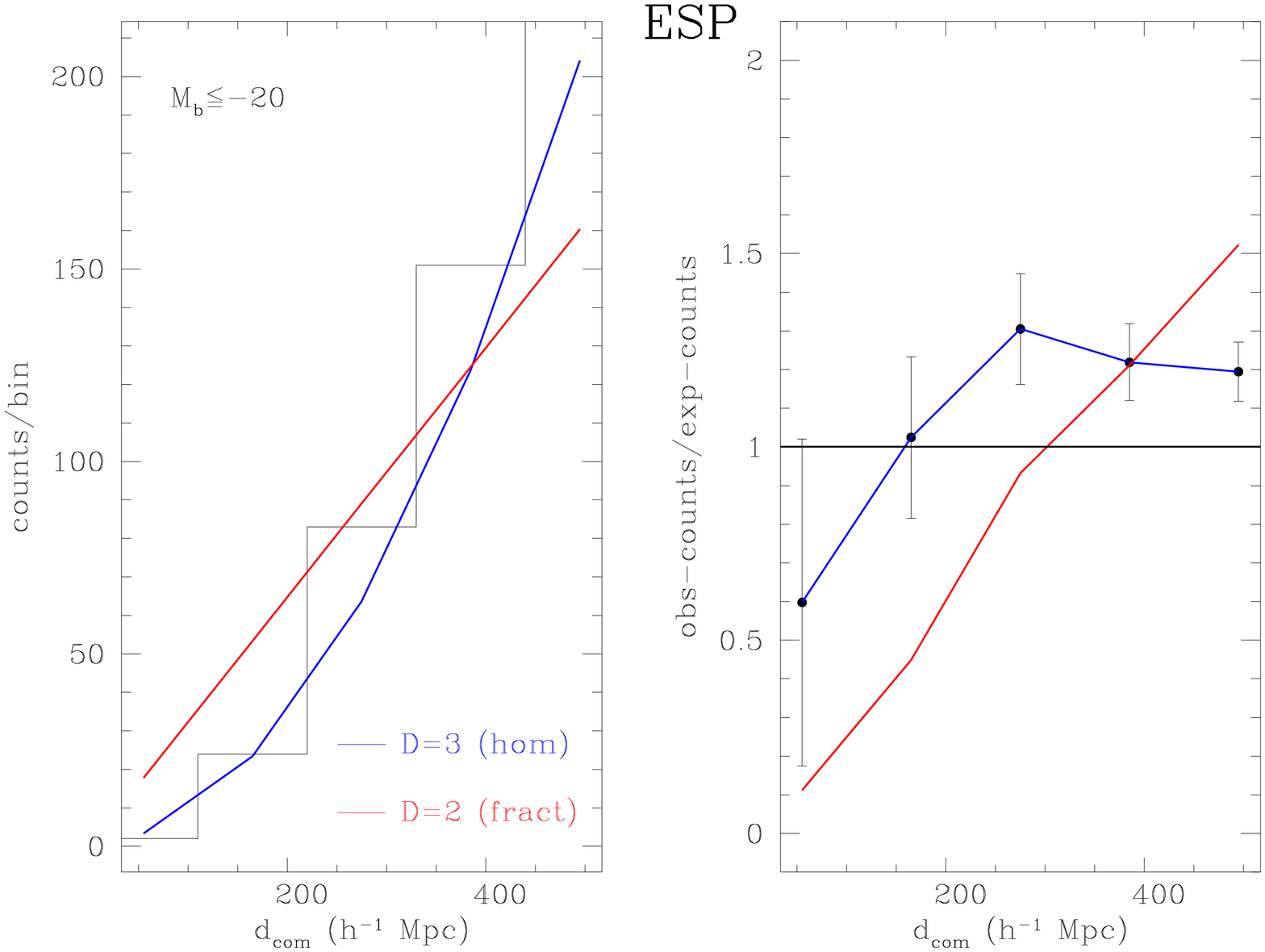,width=4.5in,angle=270}}
\caption{Differential redshift distribution for the ESP survey 
volume--limited subsample.  The plotted points and curves in the two
panels are as in figure~\ref{hist_pp20}.}
\label{esp_hist}
\endf
The important result in Figure~\ref{Nz_esp} is that the counts clearly 
stabilize around the homogeneous curve for $z>0.1$.  This is in contradiction
with what has been claimed by Pietronero and collaborators from the analysis 
of the same survey.  The only explanation of the discrepancy I can find is 
possibly in them having improperly analysed
a preliminary, incomplete version of the survey and most importantly not 
having performed any
K--correction\footnote{In fact, in a recent calculation Scaramella \etal
(1997), show explicitly that if one uses $cz/H_\circ$ as a measure of
distance and applies no K--correction, the resulting $N(cz/H_\circ)$ behaves 
similarly to a $D\sim 2$ model.}.

In Figure~\ref{esp_hist}, just as a further indication, I show the behaviour
of the differential redshift distribution for the same sample.  Again, the
$D=2$ fractal model (red line), overestimates the expected distribution at
small distances while underestimating it at large distances.

\section{So the Universe is not Fractal?} \label{true-scaling}

So far we have concentrated on two specific observational tests that I 
considered as crucial, and that explicitly show the difficulty of describing 
the distribution of galaxies as a pure unlimited fractal.  To shortly
summarize the conclusions reached so far: 1) 
the mean density and hence the correlation length does not depend on 
the sample size {\it for the largest statistically complete redshift samples
available today}, and on the contrary $r_\circ$ 
is remarkably stable among surveys of very different geometry.  2) The growth
of the number of galaxies in volumes of increasing radius is more consistent,
{\it within sufficiently large volumes}, with a homogeneous distribution than 
with a fractal.

Having established this, however, what can be said about the fractal 
properties of the galaxy distribution in the range where galaxies are 
clustered, and in which their distribution shows a clear self--similar
character?  The classical arguments in this respect are discussed in 
detail in Peebles (1980; 1993), and in Mandelbrot (1982).   
The essence of this is that galaxy clustering on scales smaller than 
$\sim 5 \hmpc$ is well described by a power--law correlation function
$\xi(r) = (r/r_\circ)^{-\gamma}$ with $\gamma \simeq 1.8$.  On these
scales $\xi(r)\gg 1$, and we have seen in \S~\ref{basics} how  
this implies a fractal behaviour characterized by a fractal dimension 
$D=3-\gamma=1.2$. N--body
simulations (e.g. Valdarnini, Borgani \& Provenzale 1992), statistical
(Peebles 1980), and thermodynamical arguments (Saslaw 1984), seem to agree
in indicating that a $\sim r^{-2}$ power--law correlation function is 
a sort of equilibrium attractor of fully nonlinear gravitational evolution.
Independently on the way we describe it, therefore, this regime
seems to be undoubtedly the product of the scale--free action of gravity.

We have also seen that, in general, a fractal regime 
is strictly evidenced by a power--law behaviour of $N(r)$,
$\Gamma(r)$ or, better, $1+\xi(r)$.  If therefore, clustering has some
sort of fractal behaviour also in the linear regime (where $\xi(r)\sim 1$ or 
smaller), one will be capable to detect it only by looking at these functions.  

Motivated by these arguments, discussed in Pietronero (1987), some years 
ago Guzzo \etal\ (1991, G91 hereafter) studied 
the behaviour of $1+\xi(r)$ in the PP and CfA1 redshift 
surveys. The result was that consistently in both surveys this function 
is characterized by two 
distinct power--law ranges: a small--scale part with $D=1.2$ for $r<3.5 
\hmpc$, and a second range, between
3.5 and $\sim 20-30 \hmpc$, where $1+\xi(r)$ is still well described by
a power law, but with a fractal dimension $D\simeq 2.2$.  They
speculated that the two ranges could be evidence for two
distinct physical regimes: (1) fully nonlinear gravitational clustering on
small scales, and (2) quasi--linear fluctuations on larger scales,
characterized by a shallower clustering, possibly still reminiscent of initial 
conditions.  The interest in this latter range was clearly motivated by the 
attractive possibility to relate it directly to the linear shape 
of the power spectrum of density fluctuations $P(k)$. The observed behaviour
implied a rather steep $P(k)$ on these scales, with $n = -D \sim -2.2$, 
significantly steeper than, e.g., that expected from a standard Cold Dark
Matter (CDM) fluctuation spectrum on the same scales, that has an effective
index close to $n=-1$.  Branchini \etal\ (1994) 
showed how a phenomenological linear power spectrum with such a slope
could be 
numerically evolved to develop the correct small--scale $D=1.2$ slope, 
while providing at the same time a good match to the observed large--scale
shallower clustering.

The result by G91 was certainly approximate in many respects, in
particular in the treatment of redshift--space distortions.
The two regimes can indeed be clearly seen
only in real space: redshift--space damping produced by cluster
velocity dispersions below $2-3 \hmpc$ has the perverse effect of 
flattening down the correlation function, bringing the observed
$D$ to a value around 1.6--1.8, so that when one plots $1+\xi(s)$,
this is very well described by a single power law with slope $D-3 \sim -1$
on all scales out to $20-30 \hmpc$.   For this reason, the best test of the
G91 findings is provided by the distortion--free clustering
measures computed from angular surveys like the APM.  Qualitatively, 
one could already
notice that the angular correlation function $w(\theta)$ from the APM 
survey  (Maddox \etal\ 1990), shows a ``shoulder'' above $1^\circ$
($\sim 4\hmpc$ at the depth of the Lick catalogue, to which that result is 
scaled), the same feature that G91 showed to be the fingerprint 
of the large--scale shallower 
scaling (see also Calzetti \etal 1992 for a similar inference from earlier
angular data).  This indication was made more 
explicit by the estimate of the APM real--space power spectrum, that Baugh 
\& Efstathiou (1993), computed by de--projecting $w(\theta)$.
I have replotted their estimate in Figure~\ref{pk_APM}.
\begf
\centerline{\psfig{figure=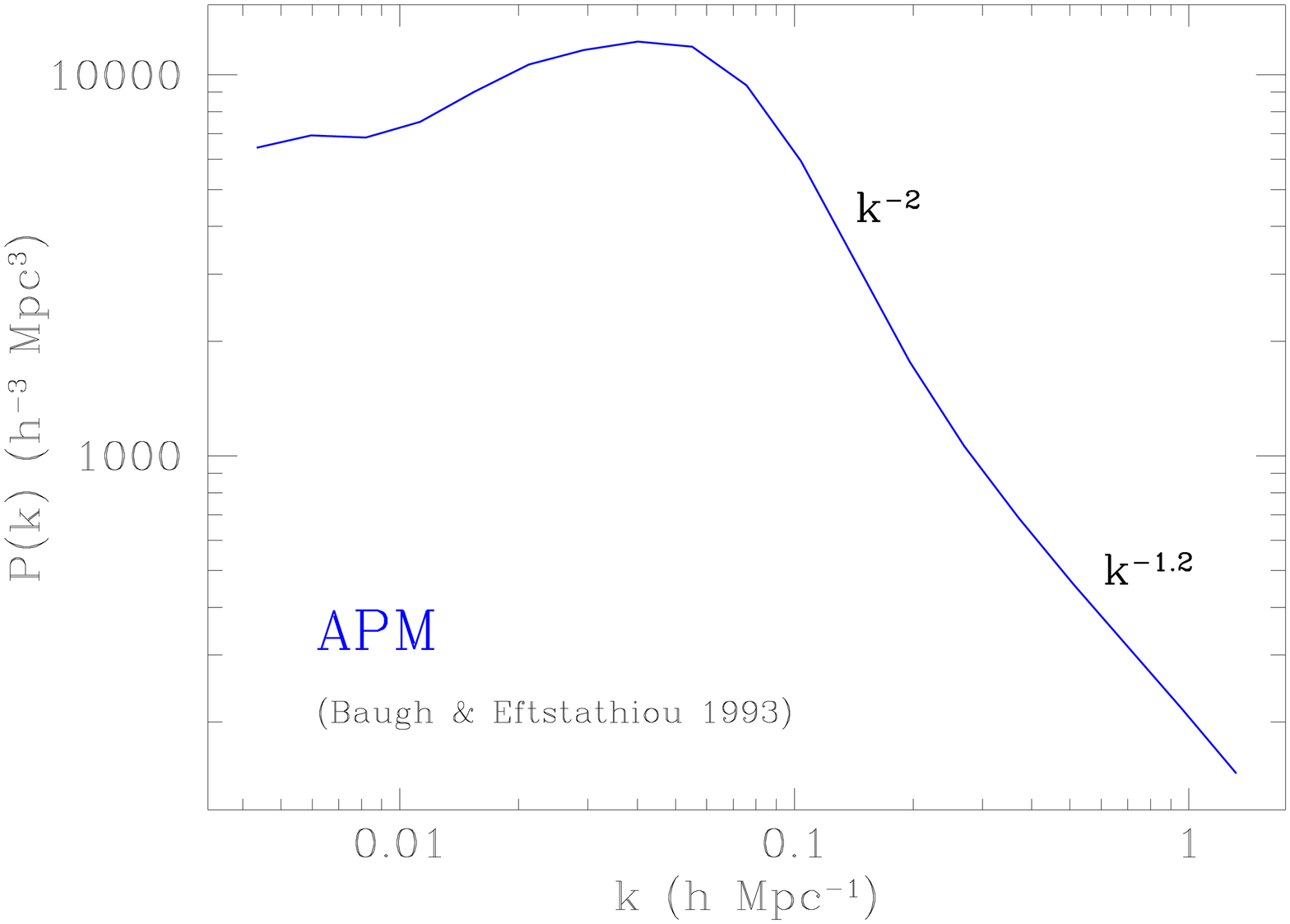,width=5in,angle=270}}
\caption{The real--space power spectrum measured from the APM galaxy survey.
Note the two power--law regimes for $k > 0.08$ h Mpc$^{-1}$, that 
correspond in Fourier space to 
the two similar scaling ranges observed in $1+\xi(r)$.  Note how the 
logarithmic slope $n$ of the power spectrum is directly related to the 
fractal correlation dimension $D=-n$ in the corresponding range of scales.
Error bars are omitted for clarity (see Baugh \& Efstathiou 1993).}
\label{pk_APM}
\endf
\begf
\centerline{\psfig{figure=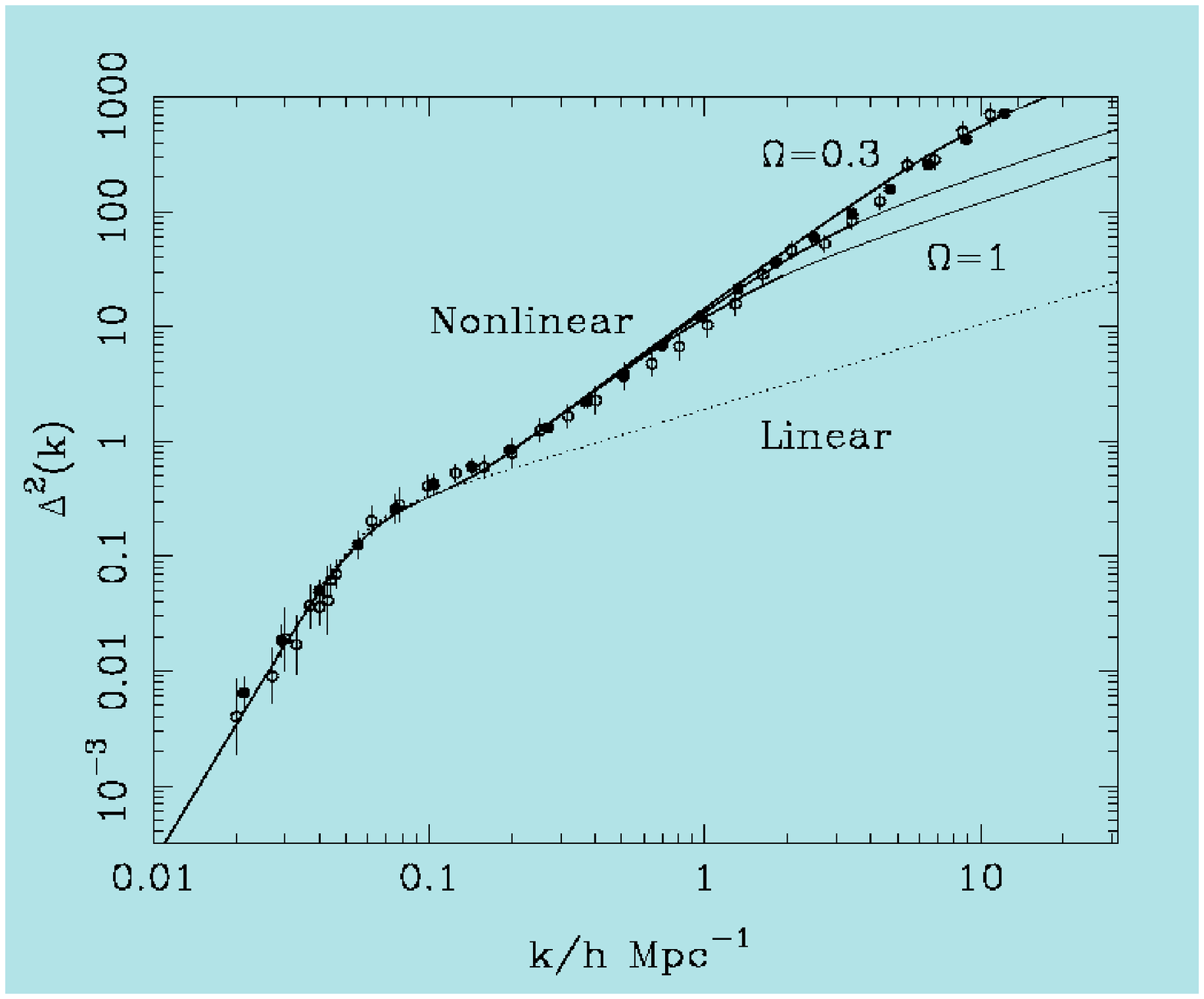,width=5in}}
\caption{Observed power spectra from the APM (filled circles), 
and \IRAS--{\it QDOT} (open circles) surveys, from Peacock (1997).  The dotted
line is the phenomenological model for the linear $P(k)$ that
best reproduces the observed data, characterized by a power--law index 
above 0.08 h Mpc$^{-1}$ $n \sim -2.2$.  Power spectra
are here plotted as the logarithmic contribution to the variance,
$\Delta^2(k) = {\rm d}\sigma^2/{\rm d}\ln{k} \propto k^3 P(k)$.  
[Reproduced with permission from the MNRAS].
}
\label{jap}
\endf
The figure shows that at wavenumbers larger than the 
peak around 0.05 h Mpc$^{-1}$ the spectrum
is characterized by two slopes: $k^{-2}$ at intermediate wavenumbers, and
$k^{-1.2}$ at larger $k$'s (small spatial scales).  These correspond,
in Fourier space, to the two clustering ranges observed by G91 on 
$1+\xi(r)$.  The transition between the two regimes is around 
$k\simeq 0.2$ h Mpc$^{-1}$, corresponding to the
$r \sim 3-4 \hmpc$ change of slope observed in the correlation function
[$r\sim \lambda/4 = \pi/(2k)$]. 
This scale, therefore, seems to assume a particularly important dynamical
meaning in the clustering properties of our Universe at the present epoch:
it is the typical scale below which {\it today} fluctuations are 
fully nonlinear.

The relation of the two power--law ranges to the linear and nonlinear 
regimes of evolution of the power spectrum, has been recently confirmed
through a detailed investigation by Peacock (1997).   Branchini \etal\ (1994)
simply assumed that the large--scale $D\simeq 2.2$ power law was strictly
describing the behaviour of linear fluctuations, and extrapolated this 
regime to smaller scales to construct their phenomenological linear 
power spectrum.  Peacock (1997), on the other hand, applies a modified 
version of Hamilton \etal\ (1991) analytical reconstruction
algorithm to the Baugh \& Efstathiou APM power spectrum and 
to a similar real--space estimate from the \IRAS--{\it QDOT} survey (Saunders \etal\
1992), to obtain the linear power spectrum from the nonlinear data.  
The observations [expressed as 
variance per logarithm of $k$, $\Delta^2(k) \propto k^3 P(k)$],
and the reconstructed linear power spectrum (dotted line), are shown 
in Figure~\ref{jap}.   The important result is that the reconstructed linear
power spectrum turns out to be characterized by a an index around $n=-2.2$ 
for scales smaller than the turnover at $k\sim 0.08$ h Mpc$^{-1}$, 
as deduced by G91 and Branchini \etal\ (1994) using the shape of $1+\xi(r)$.

\section{Discussion and Conclusions}

The main conclusions that I have reached in this harangue in defense of
a homogeneous Universe can be summarized as follows.

1) There is strong evidence from the large--scale distribution of galaxies
that the observed clustering does not continue to arbitrarily large scales,
but is limited by a turnover to homogeneity.  This cutoff scale, that we can
associate to a maximum in the power spectrum $P(k)$, is most
probably located around a 3D wavelength scale comprised between 100 and $200 
\hmpc$ and becomes evident 
when we properly analyze the largest surveys presently available.
It is clear that even these samples have just reached
this scale, and certainly we are not yet sampling well enough the largest
clustering wavelengths in the galaxy density field.  However, the 
stability of the galaxy--galaxy correlation length and the law of 
growth of the number of objects within larger and larger volumes are 
in contraddiction with a pure fractal with a well--defined $D$ on all
observed scales.  

2) On the other hand, within the range where they are clustered galaxies 
do cluster in a fractal--like way.  More specifically, there is strong
evidence that we are witnessing the effect of 
two processes, both producing a scale--free behaviour in the 
galaxy distribution within their respective ranges of action.  
At separations smaller than $\sim 3.5 \hmpc$ we observe a distribution 
with $D=1.2$, resulting from the nonlinear 
gravitational evolution of the initial spectrum.   On larger scales,
there is a second range extending to separations of $20-30 \hmpc$
in which a fractal--like scaling is 
also observed, but with a shallower scaling law characterized by 
$D = 2 - 2.3$.
We have seen that there are very good reasons to interpret this range 
as being directly related to initial conditions, corresponding to a
linear power spectrum with a slope $n$ between $-2$ and $-2.3$.  In the
framework of CDM models, open models with $\Omega\simeq 0.3$ 
best reproduce the observed shape on small and large scales (Peacock 1997).  
However,
the best match to the observations in absolute terms is provided by 
phenomenological shapes in which the maximum of $P(k)$
is much sharper than in any CDM variant (Branchini \etal\ 1994; 
Peacock 1997)\footnote{Although it needs to be confirmed by better 
large--scale data, see also the direct evidence for a sharp turnover
in $P(k)$ recently provided by Einasto \etal\ (1997).}.

If this primordial scaling really corresponds to a fractal behaviour
(to this end it would be important a better understanding of the present 
phase distribution of the large--scale density field as traced by galaxies,
see e.g. Ghigna \etal\ 1994), 
then one can speculate how this could have developed
strarting from the natural assumption of an initially gaussian density field.
If initial conditions were gaussian
(and therefore intrinsically non--fractal),
characterized by a power--law $P(k)$ with $n\sim -2.2$ over 
comoving wavelengths between $\sim 1-100 \hmpc$ and random phases, 
one could think
of linear gravitational clustering (perhaps in combination with 
galaxy formation processes that favour the density field above 
a certain threshold -- see e.g. Szalay \& Schramm 1985), as a process
that very quickly rearranges the phases of the density field, while
the power spectrum slowly grows self--similarly in amplitude.  
This would provide the necessary phase ordering which is a key 
for producing fractal distributions,
characterized by strongly non--random phases.  In a different
perspective, this stage of evolution could be seen as the build
up of pancakes, i.e. the early linear growth of long--wavelength
fluctuations along one preferred direction, giving rise to
flattened, sheet--like structures (an interpretation already proposed by Dekel 
\& Aarseth 1984).  In fact, Provenzale \etal\ (1994) explicitly 
showed that the observed two regimes of clustering 
can be reproduced also in a toy model in which: (1) galaxies 
are distributed as a 2D fractal on the surfaces of planar 
structures; (2) these are in turn randomly oriented in space in a 3D network,
with a specific mean inter--plane distance.  In this picture, the
large--scale $D\simeq 2$ dimension would be {\it topological}, rather 
than fractal.   

While large scales grow linearly, at the same time on small scales 
nonlinear gravitational 
clustering starts building up vigorously a different, more packed 
structure with $D\sim 1.2$,
that grows up in time extending to larger and larger scales.  Is this
a true fractal distribution?   It has recently been shown explicitly
that the observed scaling law can be produced by a distribution of 
{\it density singularities} (i.e. clusters) with a certain profile
(Murante \etal\ 1997).  The observed $D=1.2$ at small separations
would directly descend in this picture from the mean 
density profile of these clusters, that dominate two--point statistics 
on small scales.

My feeling is that probably on large scales (the $D\sim 2$ regime),
we are seeing a true fractal
essentially produced both by the re--ordering of phases along large--scale
density fluctuations and the effects of galaxy formation that bias the
luminous objects towards the large--scale density peaks.  
The result of these early processes
is to form the tridimensional network of objects with scale--free clustering
properties, that we observe today on scales between 5 and $\sim 50 \hmpc$. 
On small
scales, on the other hand, I find the singularity picture more realistic.  
In fact, nonlinear
collapse (strongly anisotropic in the beginning, as driven by the initial 
long--wavelength distribution, i.e. preferentially along the 
``phase channels'' provided by the large--scale structure), 
produces a ``phase--scrambling'', with a total erasing of the initial 
information. 
The result of this process, after subsequent
re--expansion, is a virialized, relaxed structure that we call a cluster,
with a well--defined and rather isotropic density profile that plays the
major role in producing the observed small--scale $D=1.2$ slope in the 
two--point correlation function.  Given that we indeed observe these 
relaxed objects and that galaxies in clusters dominate two--point statistics
on small scales, this ``distribution of density 
singularities'', as discussed by Murante \etal\ (1997), might seem a 
better description of the observed statistical behaviour, than
that of a homogeneous fractal with the same two--point function.

At the end of the 1996 ``Critical Dialogues" meeting, L.~Pietronero and
M.~Davis agreed to wager a case of good wine (Californian vs. Italian, respectively), 
about which value for the galaxy correlation length will be measured by the Sloan
Digital Sky Survey.   The discussion I have presented here seems to point in the direction
of more Italian wine to be exported to California, rather than the opposite.   
While we wait for the
SDSS to produce its first estimate of $r_\circ$, however, we seem already able to
draw some conclusions of methodological character.  The morale of the present 
discussion could be that, like 
good wine, if fractal techniques and methods are ``taken with moderation''
they provide a useful perspective from which to interpret the nature
and evolution of large--scale structure.  This perspective is not at all
in contraddiction with the classical gravitational instability description,
given that an homogeneity turnover does seem to be there. This means that,
unlike for a pure fractal, a mean density of the Universe can be defined 
and in turn the whole concept of density fluctuations makes sense.   
On the contrary, we have seen that statistical descriptions that are 
common place in the study of fractal phenomena (see e.g. Borgani 1995 for 
an overview), and that are often very
simple (but significant) modifications of those usually applied in 
cosmology, can help in evidencing very important characteristics of 
galaxy clustering.  In particular, to show where and with which 
properties galaxies do display a scale--free behaviour of some sort.

Continuing with the same oenological analogy, I should conclude at this
point that also a good vintage wine can, if used immoderately, lead to
rather annoying consequences.  In the same way, I hope to have
shown that the beauty and elegance of the fractal description has to be 
applied to the real data keeping in mind the physics of structure
formation, and most importantly the whole range of subtle selection
biases that affect the observations.

\bigskip
\noindent{\bf Acknowledgments}

I am indebted to Jim Peebles for his contribution to the birth of this paper,
and for his encouraging comments on the ma\-nu\-script.   I thank 
Antonello Provenzale for many stimulating discussions on this
subject during the past years, and  Guido Chincarini, Marc Davis, Roberto Scaramella, 
Gianni Zamorani, Elena Zucca, and especially Peter Sch\"ucker for very useful 
comments and discussions.  I am grateful to all the ESP survey team for allowing 
me to discuss some results from the project in advance of publication, and to
Carlton Baugh for providing the APM power spectrum in electronic form.  John Peacock
is gratefully acknowledeged for his patience in our e-mail discussions on 
the evolution of clustering.  I would also like to 
acknowledge vigorous discussions with my opponents in this very stimulating 
controversy, Francesco Sylos--Labini and Luciano Pietronero.

\begin{appendix}

\section{Appendix: Practical Estimators of $\xi(r)$}

In P97 and in many previous works by the same authors, a basic
methodological cricitism is made of the standard technique
normally used to estimate $\xi(r)$ from redshift surveys
in the presence of sample boundaries
(e.g. Davis \& Peebles 1983).  The effect of the finite survey volume
is usually taken into account by 
normalizing the observed number of pairs at each separation to that
obtained from a set of random points distributed within the same
volume of the survey.   Without entering into the technical details
of the problem, here I would only like to mention the results of
two simple direct tests of the limits of this technique.  To this end,
Provenzale \etal\  (1994) constructed simulated fractal distributions
using the $\beta$--model algorithm\footnote{This is a very interesting
algorithm for producing synthetic fractals which are spatially random, with
different scaling ranges and/or an upper crossover to a random distribution.
For details on the algorithm, see Castagnoli \& Provenzale (1991).}, 
from which they extracted artificial mock 
surveys with classical ``cone--like'' shapes.   The result was that  they were
not able to reproduce the observed flattening of $1+\xi(r)$ at $r>50\hmpc$ [i.e. a break
in $\xi(r)$], 
from a pure unlimited fractal model "observed" as the real surveys.  
The observed behaviour was reproduced
only by a $\beta$--model with an upper cutoff to homogeneity above cubes
of $\sim 50 \hmpc$ side, corresponding to fluctuations with wavelength 
around $100 \hmpc$. 

The same conclusions had been reached independently by Lemson \& Sanders 
(1991), who instead used Voronoi tesselation and L\'evy flights
to produce synthetic fractals by which to test the estimators of the two--point
correlation function.

\end{appendix}

\end{document}